\documentclass{ws-ijmpe}
\usepackage{amssymb}

\begin{document}

\markboth{Hilmar Forkel}{Gauge-Invariant Soft Modes in Yang-Mills Theory}

\catchline{}{}{}{}{} 

\title{GAUGE-INVARIANT SOFT MODES IN YANG-MILLS THEORY}
\author{{\footnotesize HILMAR FORKEL}}

\address{Departamento de F\'isica, ITA-CTA, 12.228-900 S\~ao Jos\'e dos Campos, 
S\~ao
Paulo, Brazil and \\ Institut f\"ur Theoretische Physik, Universit\"at
Heidelberg, D-69120 Heidelberg, Germany}

\maketitle

\begin{history}
\received{(received date)}
\revised{(revised date)}
\end{history}

\begin{abstract}
A gauge-invariant saddle point expansion for the Yang-Mills vacuum
transition amplitude on the basis of the squeezed approximation to the
vacuum wave functional is outlined. This framework allows the identification
of gauge-invariant infrared degrees of freedom which arise as dominant sets
of gauge field orbits and provide the principal input for an essentially
analytical treatment of soft amplitudes. The analysis of the soft modes
sheds new light on how vacuum fields organize themselves into collective
excitations and yields a gauge-invariant representation of instanton and
meron effects as well as a new physical interpretation for Faddeev-Niemi
knots.
\end{abstract}

\bigskip

The essence of the soft dynamics behind the most important QCD vacuum and
hadron properties is expected to involve just a few types of gluonic
long-wavelength modes. The quest for these infrared degrees of freedom
(IRdofs) has focused primarily on semiclassical fields like instantons\cite%
{sch98}, monopoles\cite{che97,tho76} and vortices\cite{tho78} and inspired
the development of various vacuum models. All these field configurations are
gauge-dependent, however, and thereby obstruct the analytical treatment and
physical interpretation of their ensembles. Gauge-invariant or fully
gauge-fixed formulations of the dynamics\cite{mak81,chr80}, on the other
hand, generally involve nonlocalities and are therefore at least as
difficult to handle analytically. Moreover, the underlying degrees of
freedom typically receive contributions from a wide variety of gauge fields
and therefore obscure the relation to gauge-dependent IRdofs.

Below we will outline an approach\cite{for06} which circumvents the problems
of dealing with individual gauge fields. Instead, it treats the
contributions from their minimally gauge-invariant generalizations, i.e.
their gauge orbits, jointly. This preserves traceable links to the gluon
content and results in a representation by local matrix fields which gather
universal contributions from dominant orbits to soft amplitudes. These
collective gluonic IRdofs emerge as contributions to the vacuum overlap
amplitude of SU$\left( N\right) $ Yang-Mills theory\ in the Schr\"{o}dinger
representation\cite{jac90}, 
\begin{equation}
Z^{\prime }:=\left\langle 0,t_{+}|0,t_{-}\right\rangle =\int D\vec{A}\Psi
_{0}^{\ast }\left[ \vec{A},t_{+}\right] \Psi _{0}\left[ \vec{A},t_{-}\right]
.  \label{zprime}
\end{equation}%
Starting from an approximate wave functional 
\begin{equation}
\psi _{0}^{\left( G\right) }\left[ \vec{A}\right] =\exp \left[ -\frac{1}{2}%
\text{ }\int d^{3}x\int d^{3}yA_{i}^{a}\left( \vec{x}\right) G^{-1ab}\left( 
\vec{x}-\vec{y}\right) A_{i}^{b}\left( \vec{y}\right) \right]  \label{gan}
\end{equation}%
of Gaussian type, we implement asymptotic freedom (in $G^{-1}$) and obtain
the associated, gauge-invariant vacuum wave functional by integrating over
the gauge group as 
\begin{equation}
\Psi _{0}\left[ \vec{A}\right] =\sum_{n}e^{iQ\theta }\int D\mu \left(
U^{\left( Q\right) }\right) \psi _{0}\left[ \vec{A}^{U^{\left( Q\right) }}%
\right] =:\int DU\psi _{0}\left[ \vec{A}^{U}\right]  \label{ginvvwf}
\end{equation}%
($d\mu $ is the Haar measure, $Q$ the homotopy degree and $\theta $ the
vacuum angle). Merits and potential shortcomings of the Gaussian ansatz (\ref%
{gan}) are discussed in Ref.\cite{for06}.

After interchanging the order of integration over gauge fields and group,
factoring out a gauge group volume and evaluating the $\vec{A}$ integral
exactly, one ends up with a functional integral $Z=\int DU\exp \left(
-\Gamma _{b}\left[ U\right] \right) $ over the ``relative'' gauge
orientation $U\equiv U_{-}^{-1}U_{+}$ where the 3-dimensional Euclidean bare
action\cite{kog95} is 
\begin{equation}
\Gamma _{b}\left[ U\right] =\frac{1}{2g_{b}^{2}}\int d^{3}x\int
d^{3}yL_{i}^{a}\left( \vec{x}\right) D^{ab}\left( \vec{x}-\vec{y}\right)
L_{i}^{b}\left( \vec{y}\right) .  \label{effact}
\end{equation}%
The $U$-dependence enters through the Maurer-Cartan forms $L_{i}=U^{\dagger
}\partial _{i}U=:L_{i}^{a}\frac{\tau ^{a}}{2i}$ and higher-order\
corrections to the bilocal operator $D^{ab}=\left[ \left( G+G^{U}\right)
^{-1}\right] ^{ab}\simeq \frac{1}{2}G^{-1}\delta ^{ab}+...$ Hence $\Gamma
_{b}\left[ U\right] $ gathers all contributions to $Z$ whose approximate
vacua $\psi _{0}$ at $t=\pm \infty $ differ by the relative gauge
orientation $U$. The variable $U$ thus represents the contributions of a
specifically weighted gluon orbit ensemble to the vacuum overlap and is
gauge-invariant by construction.

In order to access the physics which contributes to soft Yang-Mills
amplitudes with external momenta $\left| \vec{p}_{i}\right| $ smaller than a
typical hadronic scale $\mu $, we now combine a renormalization group
evolution of the bare action - to integrate out the UV modes with momenta $%
\left| k_{i}\right| <\mu $ explicitly - with a subsequent soft gradient
expansion. The result is a local effective Lagrangian 
\begin{equation}
\mathcal{L}\left( \vec{x}\right) =-\frac{\mu }{2g^{2}\left( \mu \right) }tr%
\left[ L_{<,i}\left( \vec{x}\right) L_{<,i}\left( \vec{x}\right) +\frac{1}{%
2\mu ^{2}}\partial _{i}L_{<,j}\left( \vec{x}\right) \partial
_{i}L_{<,j}\left( \vec{x}\right) +...\right]  \label{efflagr}
\end{equation}%
($g\left( \mu \right) $ is the one-loop coupling, $L_{<,i}=U_{<}^{\dagger
}\partial _{i}U_{<}$ where $U_{<}$ contains only Fourier modes with $\left|
k_{i}\right| <\mu $, and $\mu \simeq 1.3-1.5$ GeV) which has the form of a
controlled expansion in powers of $\left( \left\| \partial _{i}U_{<}\right\|
/\mu \right) ^{2}$. For practical purposes we found the truncation at second
order to yield a sufficient approximation (at the few percent level) to the
full action. The locality and structural simplicity of the Lagrangian (\ref%
{efflagr}) is a benefit of reformulating the dynamics in terms of
gauge-invariant soft-mode fields.

The above preparations enable us devise a practicable steepest-descent\
expansion for the functional integral $Z_{<}$ over the soft modes, based on
the\ saddle points $\bar{U}_{i}\left( \vec{x}\right) $ which solve%
\begin{equation}
\left. \frac{\delta \Gamma \left[ U_{<}\right] }{\delta U_{<}\left( \vec{x}%
\right) }\right| _{U_{<}=\bar{U}_{i}^{\left( Q\right) }}=0  \label{spaeq}
\end{equation}%
at fixed $Q$. The $\bar{U}_{i}^{\left( Q\right) }\left( \vec{x}\right) $
represent the IRdofs we are looking for. Their contributions to soft
amplitudes, including e.g. gluonic Green functions, are obtained by
differentiating $Z_{<}$ with respect to suitable sources. The reliability of
the leading-order approximation increases with their action value,
parametrically enhanced by the factor $g^{-2}\left( \mu \right) \gg 1$ (for $%
\mu \gtrsim 1.3-1.5$ GeV), although systematic higher-order corrections can be
calculated from fluctuations around them. The expressions for the action and
saddle point equations in terms of $\left( \phi ,\hat{n}\right) $ with $%
U_{<}\left( \vec{x}\right) =\exp \left[ \phi \left( \vec{x}\right) \hat{n}%
^{a}\left( \vec{x}\right) \tau ^{a}/2i\right] $ and $\hat{n}^{a}\hat{n}%
^{a}=1 $ for $N=2$ are given in Ref.\cite{for06}.

Important generic properties of the IRdofs include their scale stability due
to a virial theorem and three topological quantum numbers: a winding number $%
Q\left[ U\right] $ (due to $\pi _{3}\left( S^{3}\right) =Z$), a
monopole-type degree $q_{m}\left[ \hat{n}\right] $ based on $\pi _{2}\left(
S^{2}\right) =Z$ and finally a linking number $q_{H}\left[ \hat{n}\right] $
in the Hopf bundle $\pi _{3}\left( S^{2}\right) =Z$ which classifies knot
solutions. They entail the lower action bounds of Bogomol'nyi type%
\begin{equation}
\Gamma \left[ U\right] \geq \frac{12\pi ^{2}}{g^{2}\left( \mu \right) }%
\left| Q\left[ U\right] \right| ,\text{\ \ \ \ \ }\Gamma \left[ \phi
_{k}=\left( 2k+1\right) \pi ,\hat{n}\right] \geq \frac{2^{9/2}3^{3/8}\pi ^{2}%
}{g^{2}\left( \mu \right) }\left| q_{H}\left[ \hat{n}\right] \right| ^{3/4}
\label{bb}
\end{equation}%
which ensure that contributions from saddle points in high charge sectors to
soft amplitudes can generally be neglected.

While most saddle-point solutions have to be found numerically, several
nontrivial analytical solution classes, e.g. of the type%
\begin{equation}
\bar{\phi}^{\left( \hat{n}=c\right) }\left( r\right) =c_{1}+\frac{c_{2}}{%
\sqrt{2}\mu r}\left( 1-e^{-\sqrt{2}\mu r}\right) ,\text{ \ \ \ \ }\hat{n}%
^{a}=const.
\end{equation}%
exist and further, particularly symmetric ones can be obtained by solving
simplified field equations. Those include topological soliton solutions of
hedgehog type $\hat{n}^{a}\left( \vec{x}\right) =\hat{x}^{a},$ $\phi \left( 
\vec{x}\right) =\phi ^{\left( hh\right) }\left( r\right) $ whose Lagrangian
reduces to 
\begin{equation}
\mathcal{L}^{\left( hh\right) }\left( r\right) =\frac{\pi }{g^{2}\left( \mu
\right) \mu }\left[ \frac{1}{2}\left( r\phi ^{\prime \prime }\right)
^{2}+\left( 3+\mu ^{2}r^{2}\right) \left( \phi ^{\prime }\right) ^{2}+4\mu
^{2}\left( 1-\cos \phi \right) \right] .  \label{lrad}
\end{equation}%
The hedgehog saddle points turn out to comprise mainly contributions from
regions in $A$ space around the classical solutions of the Yang-Mills
equation, i.e. (multi-) instantons and merons, and were found numerically in Ref.%
\cite{for06}. (Hedgehog solutions with a monopole-type singularity at the
origin also exist.) The one-instanton dominated solution, e.g., rather closely
resembles its Yang-Mills counterpart 
\begin{equation}
\phi _{I,YM}\left( r\right) =-\frac{2\pi r}{\sqrt{r^{2}+\rho ^{2}}}
\label{omin}
\end{equation}%
with the same relative gauge orientation and $Q=1$, but also contains
quantum fluctuations. The latter stabilize the size of our solution at $\rho
\simeq 2\mu ^{-1}$ for $\mu \simeq 1.5$ GeV, compatible with instanton
liquid model\cite{shu295} and lattice\cite{mic95} results, and resolve the
IR instabilities of classical Yang-Mills instanton gases. Our meron-type
solutions share half-integer $Q$ values and infinite action (due to
angle-dependent asymptotics) with the pointlike Yang-Mills merons but remain
nonsingular and acquire a finite size due to quantum fluctuations.
Indications for a relatively large meron entropy, potentially able to
overcome their infinite-action suppression, are encountered as well. 
Moreover, we found 
solutions which have no obvious counterparts in classical Yang-Mills theory.
One of the most intriguing classes consists of solitonic links and knots.
Those emerge from a generalization of Faddeev-Niemi theory\cite{fad70},%
\begin{equation}
\mathcal{L}^{\left( \phi _{k}\right) }\left( \vec{x}\right) =\frac{\mu }{%
g^{2}\left( \mu \right) }\left[ \left( \partial _{i}\hat{n}^{a}\right) ^{2}+%
\frac{1}{\mu ^{2}}\left( \varepsilon ^{abc}\partial _{i}\hat{n}^{b}\partial
_{j}\hat{n}^{c}\right) ^{2}+\frac{1}{2\mu ^{2}}\left( \varepsilon ^{abc}\hat{%
n}^{b}\partial ^{2}\hat{n}^{c}\right) ^{2}\right] ,  \label{ln}
\end{equation}%
which turns out to be embedded in our soft-mode Lagrangian for fields of the
form $\phi _{k}=\left( 2k+1\right) \pi $ with $U_{k}\left( \vec{x}\right)
=\left( -1\right) ^{k}i\tau ^{a}\hat{n}^{a}\left( \vec{x}\right) $. Hence
our approach provides a new dynamical framework and physical interpretation
for Faddeev-Niemi-type knot solutions as gauge-invariant IR degrees of
freedom whose underlying gluon field ensembles carry a collective Hopf
charge.

In summary, we have developed a calculational framework for the Yang-Mills
vacuum transition amplitude in the Schr\"{o}dinger representation which
reveals new, gauge-invariant infrared degrees of freedom. Some of them are
related to tunneling solutions of the classical Yang-Mills equation, i.e. to
instantons and merons, while others appear to play unprecedented roles. A
remarkable new class of IR degrees of freedom consists of Faddeev-Niemi-type
link and knot solutions, potentially related to glueballs.


\begin{thebibliography}{99}
\bibitem{sch98} T. Schaefer and E.V. Shuryak, \textit{Rev. Mod. Phys.} 
\textbf{70} (1998) 323; D.I. Diakonov, \textit{Prog. Part. Nucl. Phys.} 
\textbf{51} (2003)\textbf{\ }173. For an introduction see H. Forkel, \emph{A
Primer on Instantons in QCD}, hep-ph/0009136.

\bibitem{che97} M.N. Chernodub, M.I. Polikarpov, arXiv:hep-th/9710205; G.
Bali, arXiv:hep-ph/9809351; R. Haymaker, \textit{Phys. Rep.} \textbf{315}
(1999) 153.

\bibitem{tho76} G. 't Hooft, in High Energy Physics, edited by A. Zichichi,
Editrice Compositori, Bologna, (1976); S. Mandelstam, \textit{Phys. Rep.} 
\textbf{23C} (1976) 245; G. 't Hooft, \textit{Nucl. Phys.} \textbf{B190}
(1981), 455; A. Kronfeld, G. Schierholz and U. J. Wiese, \textit{Nucl. Phys.}
\textbf{B293} (1987) 461.

\bibitem{tho78} G. 't Hooft, \textit{Nucl. Phys.} \textbf{B138} (1978) 1; J.
Greensite, \textit{Prog. Part. Nucl. Phys.} \textbf{51} (2003) 1.

\bibitem{mak81} Yu. Makeenko and A.A. Migdal, \textit{Nucl. Phys.} \textbf{%
B188} (1981) 269; A.A. Migdal , \textit{Phys. Rep.} \textbf{102} (1983) 199;
A. Dubin and Yu. Makeenko, in \emph{At the frontier of particle physics},
vol. 4, p. 2479 Ed. M. Shifman, World Scientific, Singapore (2002).

\bibitem{chr80} N. Christ and T.D. Lee, \textit{Phys. Rev.} \textbf{D22}
(1980) 939.

\bibitem{for06} H. Forkel, \textit{Phys. Rev.} \textbf{D73} (2006) 105002.

\bibitem{jac90} R. Jackiw, in ``\emph{Field theory and particle physics}'',
Eds. O.J.P. Eboli, M. Gomes, A. Santoro, World Scientific, Singapore 1990,
p. 731.

\bibitem{kog95} I.I. Kogan and A. Kovner, \textit{Phys. Rev.} \textbf{D52}
(1995) 3719.

\bibitem{shu295} E.V. Shuryak and J.J.M. Verbaarschoot, \textit{Phys. Rev.} 
\textbf{D52} (1995) 295.

\bibitem{mic95} C. Michael and P.S. Spencer, \textit{Phys. Rev.} \textbf{D52}
(1995) 4691; T. DeGrand, A. Hasenfratz and T.G. Kovacs, \textit{Nucl. Phys.} 
\textbf{B505} (1997) 417; Ph. de Forcrand, M. Garc\i a Perez and I.-O.
Stamatescu, \textit{Nucl. Phys.} \textbf{B499} (1997) 409; M.J. Teper, 
\textit{Phys. Rev.} \textbf{D58} (1998) 014505.

\bibitem{fad70} L.D. Faddeev, \emph{Quantization of Solitons}, preprint
IAS-75-QS70; L.D. Faddeev and A.J. Niemi, \textit{Nature} \textbf{387}
(1997) 58; \textit{Phys. Rev. Lett.} \textbf{82} (1999) 1624.
\end{thebibliography}
\end{document}